\documentclass[aps,prl,superscriptaddress,twocolumn]{revtex4-1}
\usepackage{units}
\usepackage{graphicx}
\begin{document}
\title{The CRESST Dark Matter Search - Status and Perspectives}
\newcommand{\mpi}{\affiliation{Max-Planck-Institut f\"ur Physik, D-80805 M\"unchen, Germany}}
\newcommand{\coimbra}{\affiliation{Departamento de Fisica, Universidade de Coimbra, P3004 516 Coimbra, Portugal}}
\newcommand{\vienna}{\affiliation{Institut f\"ur Hochenergiephysik der \"Osterreichischen Akademie der Wissenschaften, A-1050 Wien, Austria \\ and Atominstitut, Vienna University of Technology, A-1020 Wien, Austria}}
\newcommand{\tum}{\affiliation{Physik-Department and Excellence Cluster Universe, Technische Universit\"at M\"unchen, D-85747 Garching, Germany}}
\newcommand{\tuebingen}{\affiliation{Eberhard-Karls-Universit\"at T\"ubingen, D-72076 T\"ubingen, Germany}} 
\newcommand{\oxford}{\affiliation{Department of Physics, University of Oxford, Oxford OX1 3RH, United Kingdom}}
\newcommand{\wmi}{\affiliation{Walther-Mei\ss ner-Institut f\"ur Tieftemperaturforschung, D-85748 Garching, Germany}}
\newcommand{\lngs}{\affiliation{INFN, Laboratori Nazionali del Gran Sasso, I-67010 Assergi, Italy}}

\author{F.~Reindl}
\email{florian.reindl@mpp.mpg.de}
  \mpi

\author{G.~Angloher}
  \mpi

\author{A.~Bento}
\coimbra 

\author{C.~Bucci}
\lngs 

\author{L.~Canonica}
\lngs 

\author{X.~Defay}
\tum 

\author{A.~Erb}
  \tum
  \wmi

\author{F.~v.~Feilitzsch}
\tum 

\author{N.~Ferreiro~Iachellini}
\mpi

\author{P.~Gorla}
\lngs 

\author{A.~G\"utlein}
\vienna

\author{D.~Hauff}
\mpi 

\author{J.~Jochum}
\tuebingen 

\author{M.~Kiefer}
\mpi

\author{H.~Kluck}
\vienna

\author{H.~Kraus}
  \oxford

\author{J.-C.~Lanfranchi}
\tum

\author{J.~Loebell}
\tuebingen

\author{A.~M\"unster}
\tum

\author{C.~Pagliarone}
\lngs 

\author{F.~Petricca}
\mpi 

\author{W.~Potzel}
\tum 

\author{F.~Pr\"obst}
  \mpi

\author{K.~Sch\"affner}
\lngs 

\author{J.~Schieck}
\vienna 

\author{S.~Sch\"onert}
\tum 

\author{W.~Seidel}
\mpi 

\author{L.~Stodolsky}
\mpi 

\author{C.~Strandhagen}
\tuebingen

\author{R.~Strauss}
\mpi 

\author{A.~Tanzke}
\mpi 

\author{H.H.~Trinh~Thi}
\tum 

\author{C.~T\"urko$\breve{\text{g}}$lu}
\vienna

\author{M.~Uffinger}
\tuebingen 

\author{A.~Ulrich}
\tum 

\author{I.~Usherov}
\tuebingen 

\author{M.~W\"ustrich}
\mpi 

\author{S.~Wawoczny}
\tum 

\author{M.~Willers}
\tum 

\author{A.~Z\"oller}
  \tum

\date{\today}

\begin{abstract}
  In the past years the spotlight of the search for dark matter particles widened to the low mass region, both from theoretical and experimental side. We discuss results from data obtained in 2013 with a single detector TUM40. This detector is equipped with a new upgraded holding scheme to efficiently veto backgrounds induced by surface alpha decays. This veto, the low threshold of \unit[0.6]{keV} and an unprecedented background level for CaWO$_4$ target crystals render TUM40 the detector with the best overall performance of CRESST-II phase 2 (July 2013 - August 2015). A low-threshold analysis allowed to investigate light dark matter particles (\unit[$<3$]{GeV/c$^2$}), previously not accessible for other direct detection experiments.
\end{abstract}

\maketitle

\section{Introduction}
Many astronomical observations made in the last decades provide overwhelming evidence for the existence of a large amount of dark matter in the Universe \cite{Planck2014}. Today, we know that finding dark matter particles corresponds to open a new area of physics beyond the Standard Model of particle physics. Theoretical models were for a long time centered around the WIMP (weakly interacting massive particle) paradigm, predicting particles with a mass in the range of \unit[$\sim$10]{GeV/c$^2$} to \unit[$\sim$1]{TeV/c$^2$}. However, in the last years alternative well-motivated scenarios were investigated, for example asymmetric dark matter models \cite{ReviewAsymmetricDM}. These models propose masses of \unit[$\sim$0.1]{GeV/c$^2$} to \unit[$\sim$10]{GeV/c$^2$} which is below the scale of classical WIMPs.

  The hunt for dark matter is carried out through complementary channels. Firstly, accelerators may produce dark matter particles in collisions of standard model particles. Secondly, indirect searches aim to detect annihilation signals of dark matter particles. The third channel is to directly observe interactions of dark matter particles with Standard Model particles. Up to now no convincing signal was found with any of the three search strategies.

In the first extensive physics run of CRESST-II (lasting from 2009-2011) an excess of events above the known backgrounds was observed. Interpreting this this excess as being induced by dark matter, predicted WIMPs of reasonable mass and cross section \cite{angloher_results_2012}. However, already at that time this result was incompatible with other direct dark matter searches and in mild tension with previous CRESST-II results \cite{brown_extending_2012}. Clarifying the origin of the excess was one major goal for the succeeding phase 2 (July 2013 - August 2015) . This contribution will focus on results obtained from the non-blind 2013 data taken with a single detector module named TUM40. The main analysis was published in \cite{angloher_results_2014} with additional information on the detector design and backgrounds given in \cite{strauss_detector_2015,strauss_betagamma_2015}.

\section{Working Principle of a Detector}
 
  CRESST-II detector modules consist of two separate detectors to simultaneously record the phonon signal and the scintillation light of a particle interaction. We define the light yield as the ratio of the energy (in electron equivalent) measured with the light detector to the energy measured with the phonon detector. We normalize the light yield of electron recoils induced by betas and gammas to one \footnote{More precisely, the light yield is normalized to one at \unit[122]{keV}, the energy of the $^{57}$Co-calibration source.}, as this event class exhibits the highest light output.

  \begin{figure}[t]
    \centering
    \includegraphics[width=\columnwidth]{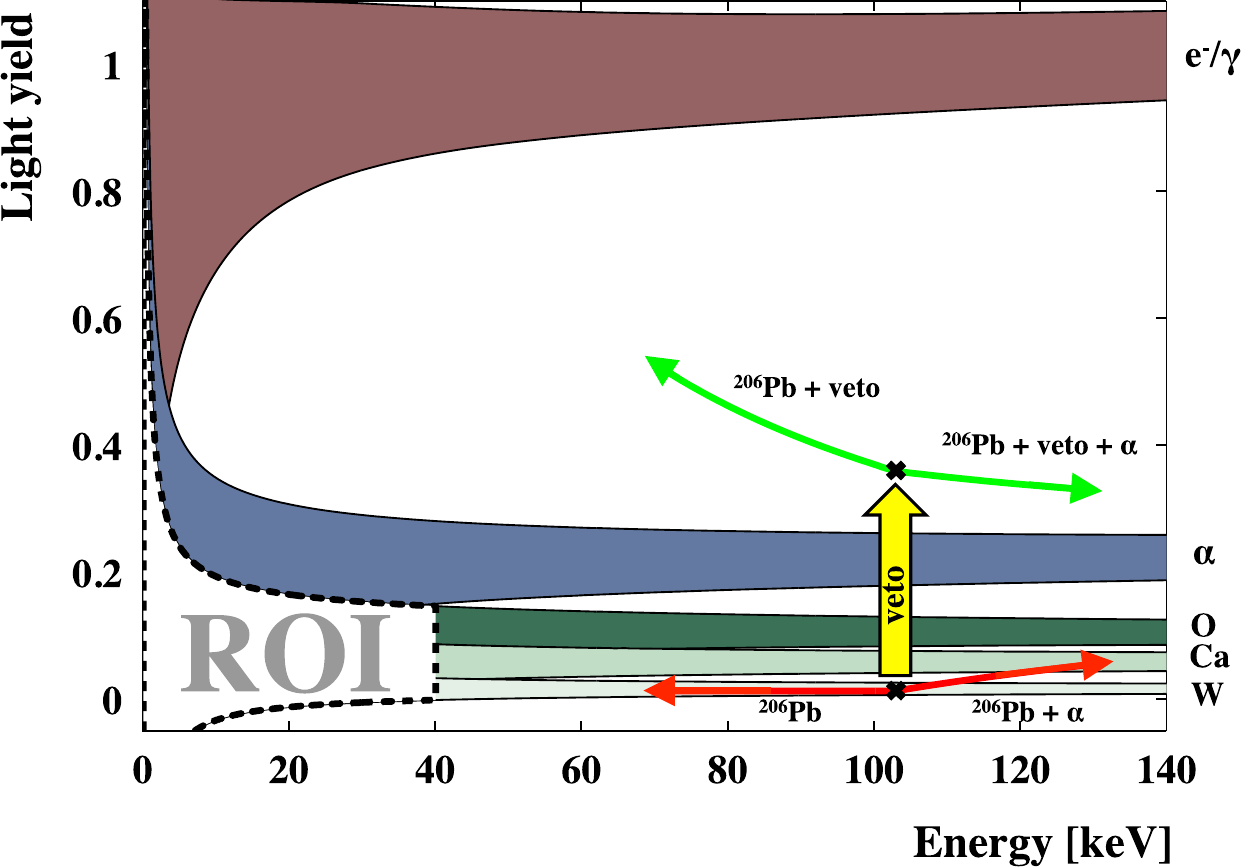}
    \caption{Light yield (light signal/phonon signal) as a function of deposited energy. The bands correspond to the different event classes, marked on the right. The region of interest (ROI) for scattering of dark matter particles includes all three nuclear recoil bands from threshold energy up to \unit[40]{keV}. The red arrows correspond to events induced by the decay of $^{210}$Po in a detector with non-active surfaces. The green errors illustrate the vetoing effect of the alpha producing light in a scintillating surface surrounding the crystal (see text). Illustration taken from \cite{strauss_detector_2015}. }
    \label{fig:LYSchematic}
  \end{figure}
  
  Figure \ref{fig:LYSchematic} depicts the light yield as a function of the energy deposited. Each band defines a region where \unit[80]{\%} of the events of the respective class (indicated on the right) are expected. The width of the bands, and thus the discrimination power between the dominant electron recoil background and the sought-for nuclear recoils, is mainly given by the resolution of the light detector. Also marked in figure \ref{fig:LYSchematic} is the region of interest (ROI) for the dark matter search. It includes scatterings off oxygen, calcium and tungsten from the energy threshold up to \unit[40]{keV}. Above \unit[40]{keV} the expected count rate for dark matter particle scatterings drops dramatically, due to the form factor \cite{Lewin1996_reviewmathematics}. The latter describes the influence of the nuclear substructure causing a deviation from the quadratic dependence of the scattering cross section on the atomic mass. This A$^2$-scaling is expected for particles scattering elastically and coherently off nuclei.

\section{Surface Backgrounds}

All classes of background events discussed in the following arise from the $\alpha$-decay of $^{210}$Po to $^{206}$Pb with an energy of \unit[5.2]{MeV} for the alpha and an energy of \unit[103]{keV} for the $^{206}$Pb recoil. Depending where this decay takes place different events may be seen in the detector. We distinguish the following cases:

\subsection{I. The $\alpha$-particle hits the crystal - degraded alpha}

A decay in the bulk of a non-scintillating material surrounding the crystal might result in a so-called degraded alpha event. The term degraded refers to the fact that the alpha may loose part of its energy before hitting the crystal. As one can see in figure \ref{fig:LYSchematic}, the alpha band overlaps with the ROI at low energies, thus representing a background for the dark matter search.

\subsection{II. The $^{206}$Pb-daughter hits the crystal - surface alpha backgrounds}

\begin{figure}[t]
  \centering
  \includegraphics[width=\columnwidth]{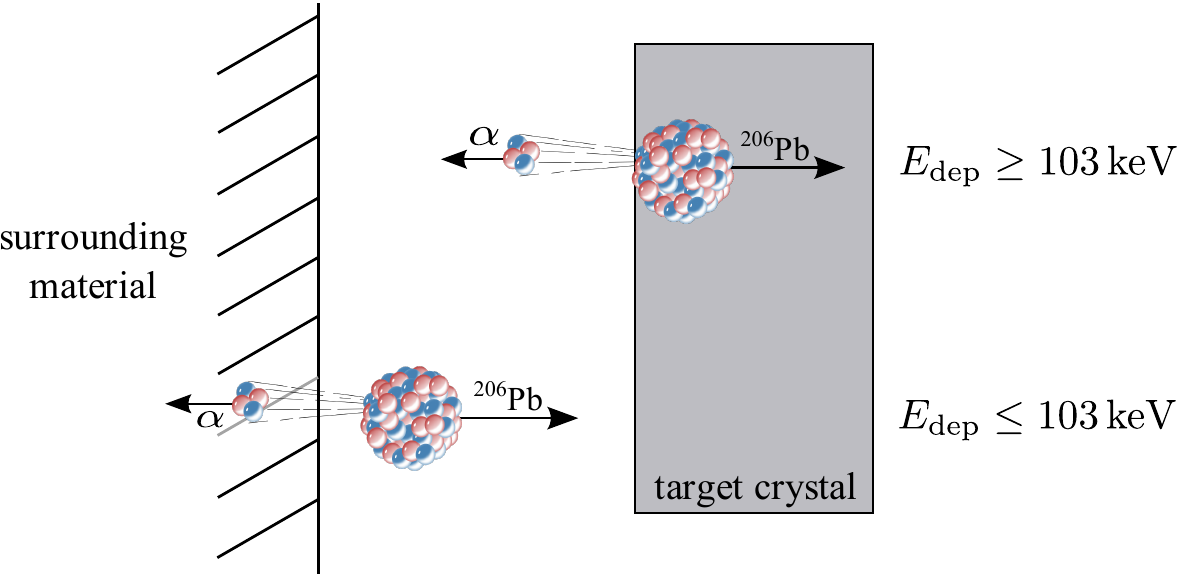}
  \caption{Illustration of the energy depositions in the target crystal induced by $\alpha$-decays of $^{210}$Po. The upper situation corresponds to a decay on the surface of the crystal, where at minimum the full energy of the $^{206}$Pb-daughter (\unit[103]{keV}) is deposited in the crystal. For a decay taking place on or very near the surface of material surrounding the crystal, \unit[103]{keV} is the maximal deposited energy (lower situation). Illustration taken from \cite{angloher_results_2012}.}
  \label{fig:SurfaceAlpha}
\end{figure}

A different situation appears for decays taking place on the surface (or on layers very close to the surface) of the crystal or any material with a line-of-sight to the crystal (illustrated in figure \ref{fig:SurfaceAlpha}). In this case the $^{206}$Pb-daughter might hit the crystal. Such lead events exhibit a light yield similar to scatterings off tungsten and, thus, potentially mimic dark matter signals. 

\subsubsection{II.a $\alpha$-decay on the surface of the crystal}
Decays on the surface of the crystal are harmless for the dark matter search as the full energy (\unit[103]{keV}) of the $^{206}$Pb-daughter is deposited in the crystal and \unit[103]{keV} is far above the region of interest. For events slightly below the surface the alpha also deposits some energy in the crystal and produces scintillation light. This event class, corresponding to the upper half of figure \ref{fig:SurfaceAlpha}, results in events indicated by the right red arrow in figure \ref{fig:LYSchematic}. 

\subsubsection{II.b $\alpha$-decay on the surface of surrounding material}
The lower part of figure \ref{fig:SurfaceAlpha} depicts alpha-decays taking place on some surface of material surrounding the crystal. In this case the $^{206}$Pb-daughter can loose part of its energy in this material and then deposits an energy of less than \unit[103]{keV}. As indicated by the left red arrow in the light-yield energy plane this class of events leaks into the region of interest.

\subsubsection{Conclusion}
In \cite{angloher_results_2012}, the leakage of lead recoil events into the region of interest was extrapolated from the observed energy spectrum of lead recoils above this region. Lead recoils were found to be the largest background source, but also for degraded alphas a substantial contribution to the number of expected background events was estimated following an analogous approach. However, as the extrapolations rely on certain assumptions to model the energy dependence of the respective event class down to the energies of interest, some uncertainty remains.

\section{Veto of Surface Backgrounds with the Upgraded Detector Module}

\begin{figure}[t]
  \centering
  \includegraphics[width=\columnwidth]{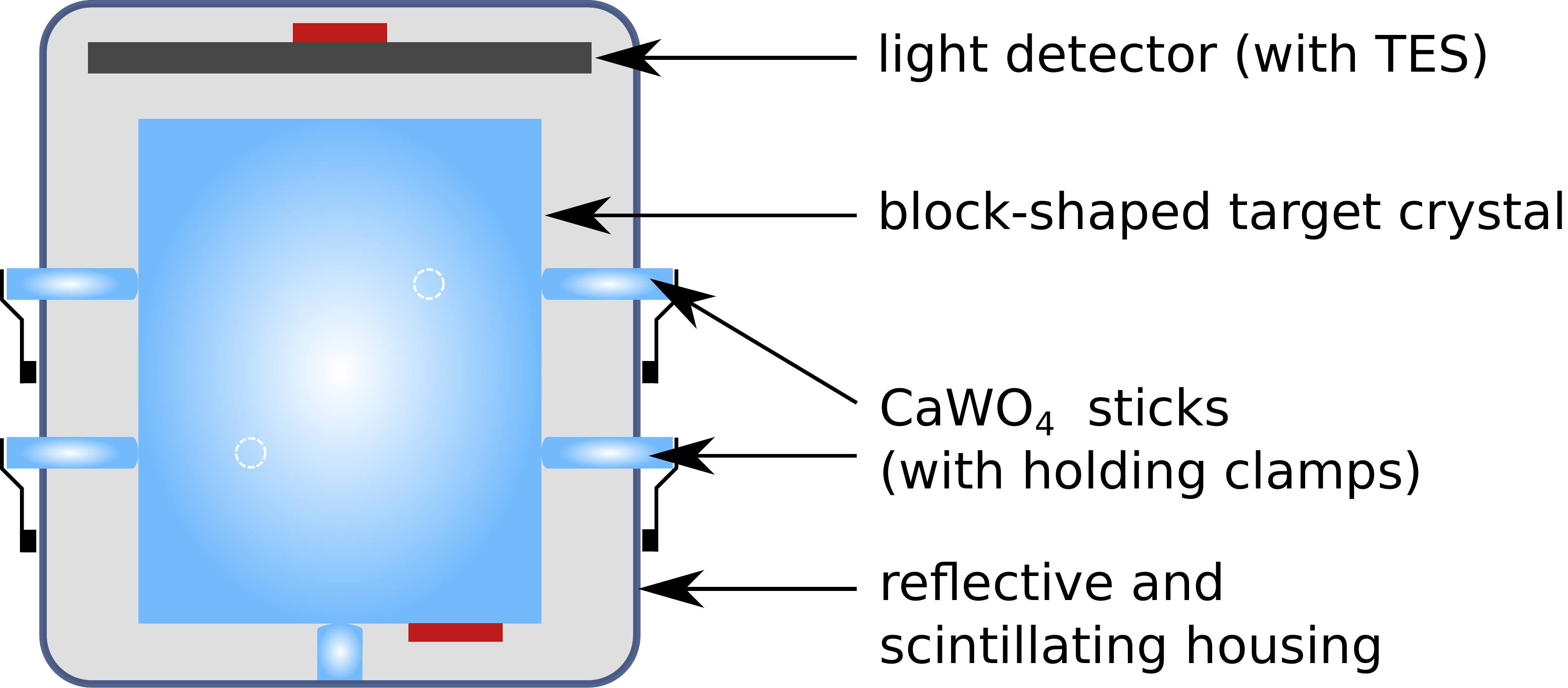}
  \caption{Scheme of the upgraded detector design using a block-shaped crystal held by CaWO$_4$ sticks. This design avoids any non-scintillating surface in the vicinity oft the crystal and, thus, vetoes events induced by surface alpha decays.}
  \label{fig:DetectorScheme}
\end{figure}

In order to eliminate the backgrounds discussed in the last section also upgraded detector modules capable to veto these backgrounds were installed in the current phase 2 \cite{reindl_status_2014}. The module under consideration - TUM40 - is one of these modules with an upgraded design. By replacing the previously used metal clamps to hold the crystal with CaWO$_4$-sticks any non-scintillating surface in the line-of-sight to the crystal has been avoided (see figure \ref{fig:DetectorScheme} for a schematic drawing). Thus, the alpha of the $^{210}$Po-decay will always produce scintillation light. This additional scintillation light lifts these events in the light yield-energy plane indicated by the yellow arrow in figure \ref{fig:LYSchematic}. This effect leads to an extremely efficient veto for the lead recoil events, as has been shown in \cite{strauss_detector_2015}. 

\section{Low-Threshold Analysis}

TUM40 was chosen for the first analysis of phase 2 as it is the module with the best overall performance. In addition to the veto for surface alpha decays, TUM40 is equipped with a crystal grown at the Technische Universit\"at M\"unchen (TUM) exhibiting an overall background level of \unit[3.51]{counts/(keV kg day)} in the energy range from 1 to \unit[40]{keV} \cite{strauss_betagamma_2015}. This value is the lowest one obtained so far for a CaWO$_4$-crystal operated in CRESST-II. Furthermore, the highly-performing phonon detector allows for a precise energy reconstruction with a resolution for low-energy gamma-lines of \unit[$\simeq$100]{eV} and for a low trigger threshold of \unit[0.6]{keV}. These features render TUM40 perfectly suited for a low-threshold analysis, which will be the focus of this section. 

\subsection{Trigger Threshold and Cut Efficiency} 

\subsubsection{Trigger Threshold}

\begin{figure}[t]
  \centering
  \includegraphics[width=\columnwidth]{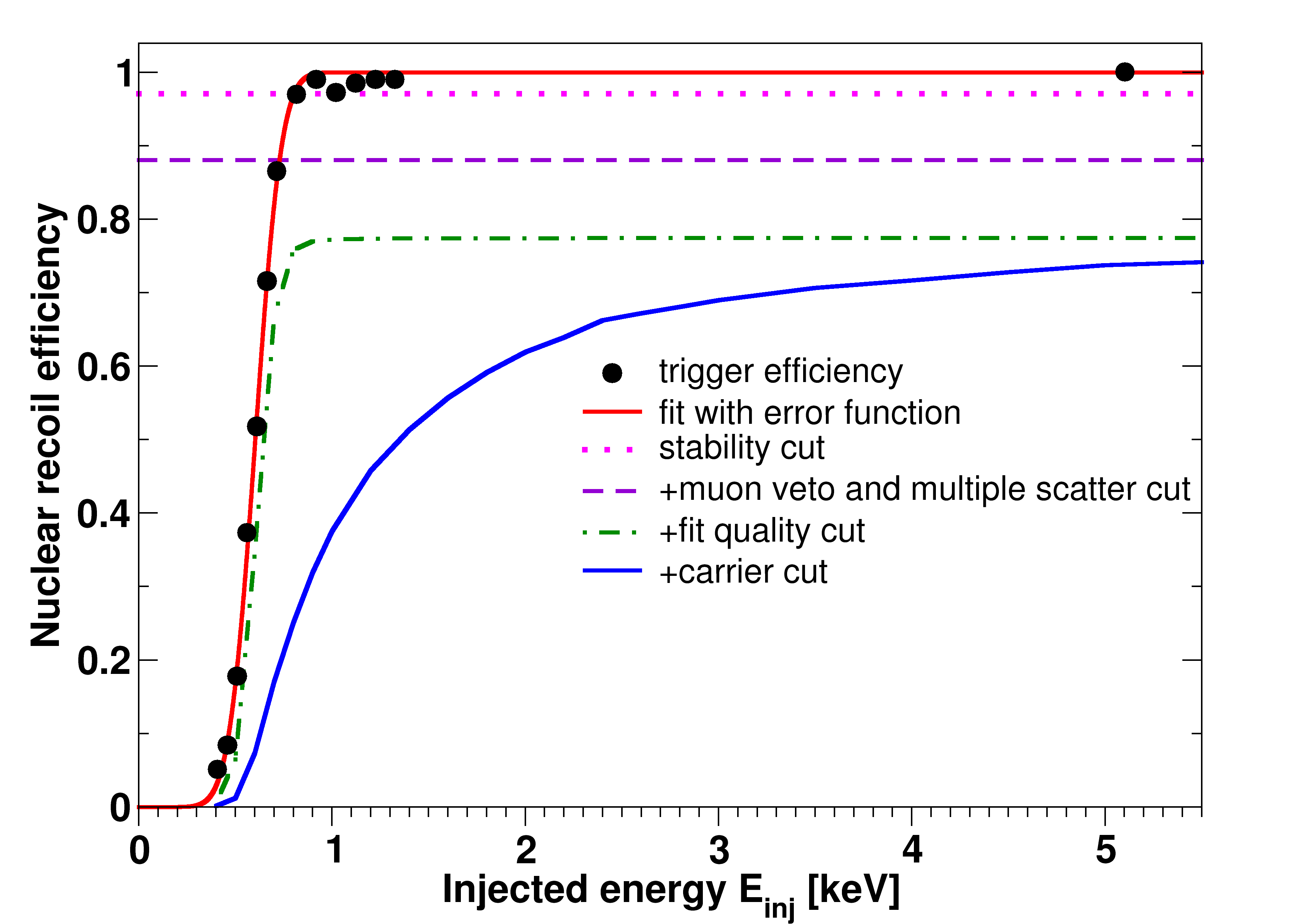}
  \caption{Trigger efficiency (black dots) as a function of the equivalent energy of the pulses injected to the heater. The red line is a fit using an error function (see text). The remaining lines show the cumulative probability for a potential signal event to survive the selection criteria as a function of equivalent energy of the simulated pulses. The blue line is the total efficiency after applying all cuts. (Plot from \cite{angloher_results_2014}.)}
  \label{fig:Eff}
\end{figure}

Obviously, a precise knowledge of the trigger threshold is a basic requirement to analyze data down to threshold. In CRESST every detector is equipped with a separated heater. Its main purpose is to stabilize the temperature in the desired operating point using a constant current. Additionally, pulses of a shape similar to those induced by particle interactions are injected. Thus, we directly measure the trigger threshold by injecting low-energy pulses to the heater and determining the fraction of pulses successfully triggered for each injected energy. The measured efficiency is depicted as a function of energy with black dots in figure \ref{fig:Eff}. The red line is the result of a fit with a function describing the convolution of a step function for an ideal trigger with a Gaussian function to account for the baseline noise of the detector. We determine the trigger threshold (\unit[50]{\%} point of the red line) to be \unit[(603$\pm$2)]{eV}.

\subsubsection{Cut Efficiency}

We apply different types of cuts on the raw data. For a low-threshold analysis energy-dependent cut efficiencies (= the probability of a potential signal event to survive all cuts) cannot be neglected. We determine the efficiencies by applying the cuts on a set of artificial events. We create these events by superimposing randomly sampled empty baselines with signal templates. As the empty baselines are affected by any artifacts in exactly the same way as potential signal events this method provides a precise estimate of the cut efficiencies.  

The first cut applied removes time periods with the phonon or the light detector (or both) not being in their stable operating point. The second cut removes coincidences between an event and the muon veto or any other detector, as multiple interactions of dark matter particles are practically excluded considering their small anticipated interaction cross section. Additionally, all events where the correctness of the energy reconstruction is in doubt have to be removed. The main tool of the energy reconstruction is a fit of the pulses with a signal template. Thus, all events where this fit yields an increased RMS value are discarded (denoted \textit{fit quality cut} in figure \ref{fig:Eff}). 

In TUM40 the TES is not evaporated directly onto the crystal, but onto a small carrier crystal which is in turn glued to the main crystal (composite design, see \cite{Kiefer2009}). Events in the TES carrier rise faster than events in the absorber and are, therefore, efficiently removed by a cut on the rise time. However, for very low energies the discrimination via the rise time becomes more difficult causing the drop in the nuclear recoil efficiency (see blue line in figure \ref{fig:Eff}).

\subsection{Data}

\begin{figure}[t]
  \centering
  \includegraphics[width=\columnwidth]{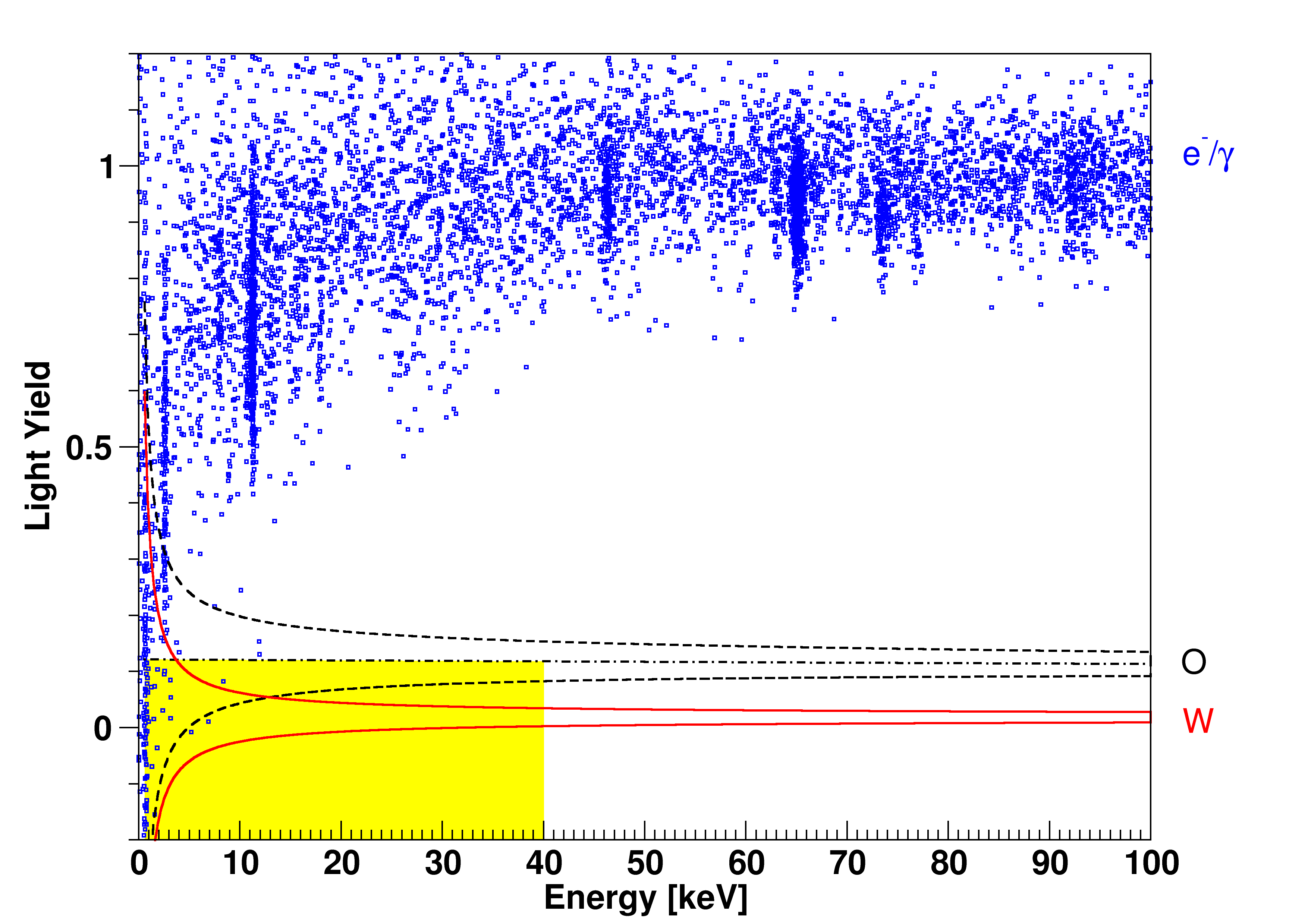}
  \caption{Data from the detector TUM40 taken in 2013 presented in the light yield-energy plane. The nuclear recoil bands (\unit[80]{\%}) for oxygen and tungsten are drawn in black and red, respectively. All events inside the yellow region are conservatively considered as potential signal events. The upper light yield boundary is set to the center of the oxygen band. (Plot from \cite{angloher_results_2014}.)}
  \label{fig:LY}
\end{figure}

In figure \ref{fig:LY} we present the light yield versus energy data of the non-blind 2013 data set taken with TUM40 corresponding to an exposure before cuts of \unit[29.35]{kg days}. Comparing the data with the schematic drawing in figure \ref{fig:LYSchematic} reveals that no background events related to surface alpha decays are observed. The gamma lines well pronounced in the e$^-$/$\gamma$-band mostly originate from cosmogenic activation of tungsten (see \cite{strauss_betagamma_2015,angloher_results_2014} for a detailed discussion). 

Because of the anticipated A$^2$-dependence of the dark matter particle scattering cross-section, scatterings off the heavy tungsten nuclei are expected to be far more numerous than on the rather light elements oxygen and calcium. However, the more the masses of two scattering partners differ, the less energy is transferred in the scattering. Thus, oxygen and calcium extend the reach towards low dark matter particles, where scatterings off tungsten are already well below threshold. 

For the argument outlined above, the region of interest in the light yield energy plane extends over all three nuclear recoil bands. However, we are using the Yellin optimum interval method \cite{Yellin2002_Limit} without background subtraction to derive an exclusion limit on the scattering cross section. Thus, by only considering events below the center of the oxygen band (yellow region in figure \ref{fig:LY}) we substantially reduce the leakage of e$^-$/$\gamma$-background events without a significant cut of the sensitivity towards low dark matter particle masses.

\subsection{Results}

\begin{figure}[t]
  \centering
  \includegraphics[width=\columnwidth]{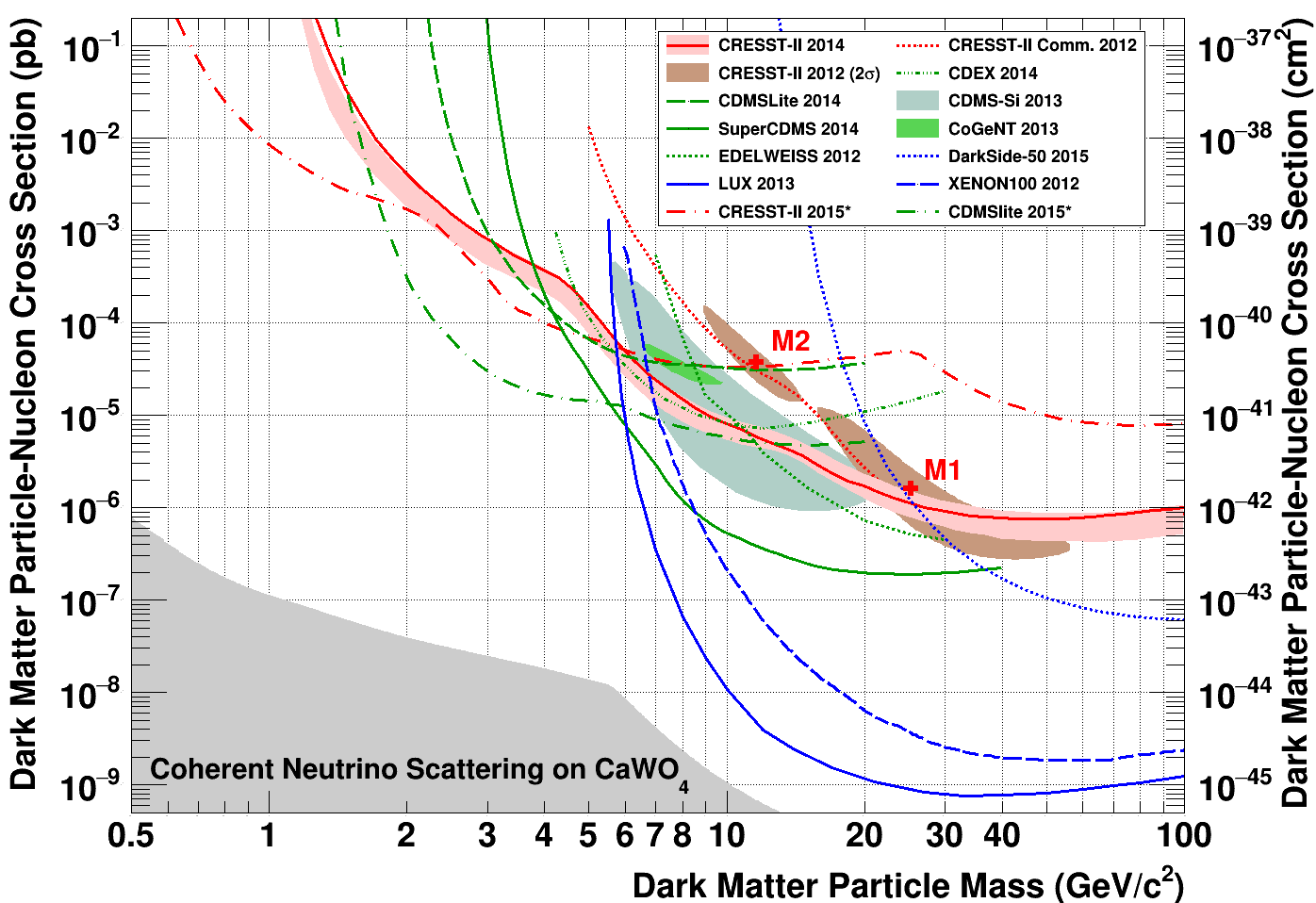}
  \caption{This plot shows the elastic, spin-independent scattering cross-section of dark matter particles with nucleons as a function of the mass of the dark matter particle. The result discussed here \cite{angloher_results_2014} is drawn in solid red, the expected statistical fluctuations as light red band. Additionally drawn are other \unit[90]{\%} exclusion limits from CRESST (red,\cite{brown_extending_2012,the_cresst_collaboration_results_2015}), germanium based experiments (green, \cite{Agnese2014,Agnese_2013CDMSlite,supercdms_collaboration_wimp-search_2015,edelweiss_collaboration_search_2012,cdex_collaboration_limits_2014}) and experiments based on liquid noble gases (blue, \cite{Aprile2012,Akerib2014,agnes_first_2015}). The gray colored region marks the limit for an CaWO$_4$-based experiment free of backgrounds induced by coherent neutrino-nucleus scattering \cite{gutlein_impact_2015}. Colored shaded areas correspond to regions of allowed parameter space reported by \cite{Aalseth2013,Agnese2013,angloher_results_2012}. Limits marked with a '*' in the legend are currently under review, but included nonetheless to show the latest advancement in the low-mass regime. }
  \label{fig:Limit}
\end{figure}

With standard assumptions on the dark matter halo we derive an exclusion limit (\unit[90]{\%} confidence level (C.L.)) on the dark matter particle-nucleon cross-section, drawn as function of the mass of the dark matter particle in solid red in figure \ref{fig:Limit} \cite{angloher_results_2014}. The light red band shows the expected statistical fluctuations (\unit[1]{$\sigma$ C.L.}), obtained by an empirical Monte Carlo simulation which is based on the events observed in the e$^-$/$\gamma$-band. The limit obtained from data meets the expectations from this model leaving no hint for events beyond the e$^-$/$\gamma$-background. 

The WIMP interpretation \cite{angloher_results_2012} of the excess observed in the previous phase 1 is to a large extent ruled-out by this result. However, the full exposure of phase 2 and a combination of several detector modules will be needed to finally clarify the origin of the excess. 

At the time of publication in July 2014 new parameter space was explored for masses below \unit[3]{GeV/c$^2$}. Very recently (September 2015), the SuperCDMS (dash-dotted green,\cite{supercdms_collaboration_wimp-search_2015}) and the CRESST (dash-dotted red, \cite{the_cresst_collaboration_results_2015}) collaborations  announced new results further pushing the sensitivity of direct dark matter searches in the low-mass regime. 

The new (2015) CRESST-II low-threshold result (\cite{the_cresst_collaboration_results_2015}) was obtained with the detector module Lise. Compared to TUM40, Lise exhibits a higher overall background level, mainly caused by a less radiopure (commercial) crystal and the accidental illumination with an $^{55}$Fe-source, installed to calibrate the light detector of an adjacent detector module. Thus, for dark matter particle masses above \unit[6]{GeV/c$^2$} the 2015 exclusion limit is significantly weaker than the 2014 limit. However, the trigger threshold of Lise is \unit[(307$\pm$3.6)]{eV}, which is almost a factor of two lower than the one of TUM40. With this low threshold we were able to open-up the direct search for dark matter particles lighter than \unit[1]{GeV/c$^2$}. 

  Both the 2014 and the 2015 exclusion limits show a more moderate rise towards lower energies compared to other direct dark matter searches. This is a consequence of the light elements present in our target material. The transition point from tungsten to oxygen being the dominant scattering partner causes the kinks in the respective exclusion limit seen at \unit[$\sim$3]{GeV/c$^2$} for the 2015 result and at \unit[$\sim$4.5]{GeV/c$^2$} for the 2014 limit.

Motivated by the growing theoretical interest in the low-mass regime and the results discussed above, the next measurement campaigns will focus on light dark matter particles, as outlined in \cite{the_cresst_collaboration_probing_2015}. The new detectors will combine (an upgraded version of) the superior TUM40 holding concept with a phonon detector specially designed towards the measurement of very small energy deposits. Details on the design of the new detectors and their potential are outlined in \cite{the_cresst_collaboration_probing_2015,strauss_procLTD}.

The analysis presented in \cite{angloher_results_2014} and in the present contribution proved for the first time the performance of CRESST detectors down to thresholds well below \unit[1]{keV}. Thus, it not only contributed to the hunt of dark matter but also laid foundation for further significant progress in sensitivity - already achieved with Lise in 2015 and expected for the near future with the upgraded detector modules.

\section*{Acknowledgments}
  \footnotesize
  We are grateful to LNGS for their generous support of CRESST, in particular to Marco Guetti for his constant assistance.This work was supported by the DFG cluster of excellence: Origin and Structure of the Universe, by the Helmholtz Alliance for Astroparticle Physics, and by the BMBF: Project 05A11WOC EURECA-XENON.

\bibliographystyle{h-physrev}
\bibliography{lise}
\end{document}